\newcounter{myctr}

\documentclass{ws-acs}
\usepackage{hyperref}
\usepackage{longtable}

\begin{document}

\makeatletter
\def\@biblabel#1{[#1]}
\makeatother

\markboth{Yu. Y. Tarasevich, A. V. Danilova, O. E. Romanovskaya}{Network analysis of verbal communications  in the novel \emph{The Master and Margarita} by Bulgakov}

%
\catchline{}{}{}{}{}
%
\title{NETWORK ANALYSIS OF VERBAL COMMUNICATIONS IN THE NOVEL \emph{THE~MASTER AND MARGARITA} BY~M.~A.~BULGAKOV}

\author{YURI Y. TARASEVICH\footnote{Corresponding author.}, ANNA V. DANILOVA and OLGA E. ROMANOVSKAYA}
\address{Astrakhan State University,\\  Astrakhan, 414056, Russia,\\ tarasevich@asu.edu.ru}

%

\maketitle

\begin{history}
\received{(received date)}
\revised{(revised date)}
\end{history}

\begin{abstract}
A network analysis of the structure of verbal communications in one of the most popular Russian novels of the Soviet era \emph{The Master and Margarita} by M.~A.~Bulgakov has been carried out. The structure of the novel is complex, i.e., there is `a story within a story'. Moreover, the real-world-characters and the other-world-characters are interacting in the novel. This complex and unusual composition makes the novel especially attractive for a network analysis. In our study, only paired verbal communications (conversations) between explicitly present and acting characters have been taken into account; frontal communications, monologues, off-stage characters as well as expected connections between characters have not been taken into account. Based on a character pair verbal communication matrix, a graph has been constructed, the vertices of which are the characters of the novel, while the edges correspond to the connections between them. Taking only paired verbal communications into account leads to the result, that the character network can be described by an ordinary, rather than a directed graph. Since the activity of the conversations was out of our intended scope, the edges have been given no weights. The largest connected component of the graph consists of 76 characters.
Centralities, such as degree, betweenness, closeness, eigenvector, and assortativity coefficient were computed to characterize the network. The assortativity coefficient of the network under consideration is negative $-0.133$, i.e., the network does not demonstrate the properties of a social network. The structure of the communities in the network was also analysed. In addition to the obvious large communities~--- the characters from the Yershalaim part of the novel and the characters of the Moscow part --- the analysis also revealed a fine structure in the Moscow component. Using the analysis of centralities, a group of main characters has been detected. The central characters of the novel are Koroviev, Margarita, Bezdomny, Woland, Behemoth, Azazello, Bosoi, Warenukha, Master, and Levi Matthew.
\end{abstract}

\begin{flushright}
`and what is the use of a book,' thought Alice\\ `without pictures or conversations?'\\
Lewis Carroll, \emph{Alice's Adventures in Wonderland}\\
\medskip
`I checked\\
Harmony with algebra.'\\
Alexander Pushkin,  \emph{Mozart and Salieri}\\
\end{flushright}

\section{Introduction}
In the last decade, analysis of the structure of characters' networks has been actively applied to religious~\cite{Massey2016,Massey2019}, mythological~\cite{Choi2007,Carron2012,Carron2013,Carron2013a,Kenna2016,Holovatch2016a,Sarkanych2016,Yose2016,Yose2018,ZhitomirskyGeffet2018,Miranda2018}, literary texts~\cite{Stiller2003,Holovatch2007,Labatut2019,Hopp2020,GesseyJones2020,Fischer2021,Rahul2021}, and graphic novels~\cite{Labatut2022}. For example, social network analyses of the some characters of the \emph{Old Testament}~\cite{Massey2016,Dekker2018}, of the four canonical Gospels and the \emph{Acts of the Apostles}~\cite{Massey2019}, of ancient Greek and Roman myths~\cite{Choi2007} and of classical poems~\cite{Miranda2018}, of the \emph{Poems of Ossian}~\cite{Yose2016}, of Rabbinic literature~\cite{ZhitomirskyGeffet2018}, of Old English epic poems~\cite{Kenna2016}, of bylinas~\cite{Sarkanych2016,Sarkanych2022}, of Shakespeare's plays~\cite{Stiller2003,Masias2016}, of \emph{A Song of Ice and Fire}~\cite{GesseyJones2020}, and classical Russian literature~\cite{Fischer2021}. The character networks of biographical, legendary and fictional texts have been studied in search for marks of genre differentiation~\cite{Holanda2019}. The analysis of such networks in the \emph{Pentateuch} showed that real social networks correspond only to the connections in the \emph{Book of Genesis}, while apparent social connections in the other books demonstrate properties that are not typical of actual social networks. The analysis recovered the close relationship between the three synoptic Gospels, but \emph{The Gospel of John} is recovered as an outgroup, reflecting its divergent nature~\cite{Massey2019}. The historicity of the events described in the \emph{Iliad} and \emph{Odyssey} found additional confirmation in the results of the analysis of the structure of the social networks referenced into these works~\cite{Miranda2018}. Network analysis has provided further evidence that Ossian may not be the author of the works attributed to him~\cite{Yose2016}.
\emph{Beowulf} has been confirmed to be realistic with the exception of the protagonist, as is reinforced by independent historical studies~\cite{Kenna2016}. A structure of social connections close to those of the real world was discovered in \emph{A Song of Ice and Fire}, which, presumably, is one of the reasons for its significant popularity~\cite{GesseyJones2020}.

Russian drama is the main object of the network analysis of the works of Russian writers images system presented in the works of F.~Fischer and D.~Skorinkin, where a total of 144 plays were studied~\cite{Fischer2021}. The methodology of a characters network creation for works intended for theatrical staging is quite simple and dictated by laws of a dramatic text construction: the connection is established between the characters, who started a dialogue in one scene. This approach let us establish that Russian drama evolution shown an increase in the size and complexity of characters network structure.

F.~Fischer and D.~Skorinkin~\cite{Fischer2021} pointed out that `the analysis of social networks of novels had been developing slightly slower' and a specific methodology is needed for extracting and formalizing network data of epic texts. According to researchers, mentioning two characters in one sentence is enough to establish a relationship between them. This method of determining connections is used to analyse L.~Tolstoy's novel \emph{War and Peace}. Network visualization of the research results exhibits 119 nodes and several communities; their density varies in different parts of the novel, what the authors of the work attribute to its content peculiarities.

F.~Fischer and D.~Skorinkin's research~\cite{Fischer2021} is one of the few works devoted to the network analysis of the Russian novel. Their results evidenced that a high degree of clustering is a common feature of the graph structure of both the realistic epic of L.~Tolstoy and the modernist work of M.~Bulgakov.


Furthermore, computer analysis of the novel \emph{The Master and Margarita} by M.~A.~Bulgakov has also been carried out~\cite{Progulova2011,Smirnov2019}. The authors of the work~\cite{Smirnov2019}, when constructing a weighted network of interactions between the characters, took into account all sentences in which characters appear together; if a character addressed another, performed some actions in relation to another character, or mentioned him/her in his/her speech, then this was considered that there is a connection between those characters. The weight of the connection depends on how many sentences describe the interactions of a particular pair of characters. The network consists of only 21 characters. The work~\cite{Progulova2011} built a network of words found in the novel: two words were considered to be connected by an edge if both of them were contained in the same sentence. In addition, the entropy of the network was calculated. Since a complete network analysis of this novel has not been carried out in previous studies, the purpose of the present study is to achieve exactly that.

Mikhail Bulgakov's novel \emph{The Master and Margarita} --- a cult book for several generations of readers --- occupies an important place in the literature of the early Soviet period and belongs to the so-called `tucked literature'. Written during the tragic years of Stalin's reign over the country, it has been published only after his death. \emph{The Master and Margarita} is a grotesquely satirical description of Soviet people life during the years of repression, it addresses the topic of relationship between an artist and totalitarian power. This work develops the best traditions of Russian satirical prose of the nineteenth century and, at the same time, addresses the themes, images and motives of I.W.~Goethe, E.A.T.~Hoffmann, it reinterprets the biblical plot. M.~Bulgakov solves `eternal' philosophical issues of ontological and axiological plans while describing modernity.


The author of the novel turned to the central motive of Goethe's tragedy \emph{Faust}, i.e.,  the appearance of the devil (Woland) among people. The author weaves the stay of Woland and his retinue in Soviet Moscow with the love story of the Master and Margarita and alongside the Master's interpretation of the gospel plot. The composition of this work of art is actually quite complex since there is `a story within a story'. Within the narrative of life in Moscow in the late 20-s and 30-s of the 20th century, the novel features the story of Pontius Pilate. An unusual and `multilayered' structure of characters network is expected due to originality of the spatio-temporal organization of the novel and the combination of a quite realistic description of the relationships between the people of the Soviet era and an emphatically fantastic otherworld. Such a novel structure may be reflected by an unusual structure of its characters network. All the characters can be divided in the three groups, viz., (i) characters of the infernal world (Woland, his retinue, ghosts at the ball, mythological characters); (ii) characters of the \emph{Novel about Pilate}, written by the Master; (iii) characters of the Moscow chapters, i.e., residents of the capital. One of the research tasks in constructing the characters network is understanding the interaction principles of different character groups in the novel. Network analysis is expected to reveal the position of Woland and his retinue (trickster characters) in the system of relationships. The preliminary novel titles included an indirect nomination of Woland, which emphasizes his image importance in the novel. A special role of Woland in the novel is indicated by an epigraph taken from Goethe's tragedy \emph{Faust}: \emph{Part of that Power, not understood, Which always wills the Bad, and always works the Good}. Woland and his trickster assistants are traditionally treated as debunkers of evil and other vices of an individual and society. A network analysis may be useful to extract more detailed functions of infernal characters in relation to the characters of other groups.

The rest of the paper is constructed as follows. Section~Methods describes some technical details of our network analysis of the novel \emph{The Master and Margarita} by M.~A.~Bulgakov. Section~Results of the analysis presents our main findings. Section~Conclusion summarizes the main results, formulates the problems identified in the course of the study, and proposes an interpretation of the results.

\section{Methods}\label{sec:Methods}
A characters network is a model of the literary text. Any model cannot and should not catch everything. However, this formal model can be studied rigorously using mathematical methods (`checked harmony with algebra' according to Pushkin). Results of such a study may be treated as reliable when composition of the characters network is well-defined.
For composing character networks various methods have been proposed depending on the way how interactions between characters are determined and accounted for. There are various ways to determine the interaction (connection, link or tie) between characters, e.g., as co-appearance at least in one scene, conversations, and friendly or hostile relations~\cite{Sarkanych2022}. An interaction can be considered as (i)~joint appearance of characters, (ii)~direct verbal interactions between the characters, (iii)~one character to explicitly mention another one, (iv)~other types of direct interaction (fighting, kissing, etc.), (v)~explicitly expressed affiliations (relatives, co-workers, etc.)~\cite{Labatut2019}.  For instance, `characters are deemed to have interacted if they directly meet each other or it is explicitly clear from the text they knew one another, even if one or both are dead by that point in the story'~\cite{GesseyJones2020}; the authors of the study devoted to a graphic novel used `interaction' in a very broad sense and manually determined, which characters are interacting based on their own understanding of the situation~\cite{Labatut2022}. It should be borne in mind that different definitions of the connections yield different insights~\cite{Everton2022}.

A network analysis of the structure of verbal communications in one of the most popular novels of the Soviet era, M.~A.~Bulgakov's \emph{Master and Margarita}, was carried out. In our study, only connections (paired verbal communications) between explicitly present and acting characters, as well as between characters, some of whose actions are explicit, and some mentioned only in someone else's story, were considered. Off-stage characters, that is, those who are only mentioned but never appear on the stage, were not taken into account. The requirement for the presence of a character on the stage automatically excludes from consideration any mentioned characters, for example, Caesar, Enanta, Niza's husband, and the former tenants of apartment~50.

In our study, connections between characters were understood as paired verbal communications, that is, dialogues. This approach is certainly not our invention. Focus on dialogues has been used previously (see, e.g., \cite{Lee2019,Pojoga2020,Lee2021}). Conversation-based network is an example of a well-defined characters network. In our study, frontal communications and monologues were not taken into account. The frontal communications were excluded as, in some cases, they are addressed to an indefinite circle of people. For example, it is impossible to determine the crowd of people whom Pilate addresses on the square, the group of people in Griboyedov who listen to Bezdomny catching the consultant, the audience of people involved in the session of black magic in the Variety, or the group of people participating in the adventures of Koroviev and Behemoth in Torgsin. Such a choice of method for determining the connections led, for example, to the  MASSOLIT management falling out of the network. However, the characters of Bosoi's dream were considered as actors. Our preliminary analysis showed that a different way of choosing connections led to a set of central characters that was difficult to interpret, and that differed from the intuitive one~\cite{Danilova2021arxiv}. Thus, when frontal communications are taking into account, Zheldybin, contrary to intuition, is one of the central characters~\cite{Danilova2021arxiv}.

Based on the character interaction matrix, a graph has been constructed, the vertices of which are the characters of the novel, while the edges correspond to the connections between them. Interaction, considered as paired verbal communication, automatically assumes that the characters network is described by an undirected graph. The edges of the graph have been given no weights, since the intensity of communications was not taken into account. Moreover, since `determining tie strength involves an element of subjectivity'~\cite{Everton2022}, we tried to avoid any subjectivity. To analyse the resulting network, the R~\cite{Rmanual} tool with the igraph~\cite{igraph} library was used.

\emph{Degree (or valency)} ($\deg V$) is the number of edges incident to a node (vertex) $V$.
In a network, a situation is possible, when nodes with a high degree are mainly connected directly with other nodes with a high degree. Such networks are called \emph{assortative}. The opposite situation is also possible, viz., nodes with a high degree are connected to other nodes with a high degree through chains of nodes that have a small number of neighbours. Such networks are called \emph{disassortative}~\cite{Newman2002}.
%
To characterize this property, the assortativity coefficient, $r$, is used. The assortativity coefficient is the Pearson correlation coefficient of degree between pairs of linked nodes. Positive values of $r$ indicate a correlation between nodes of similar degree, while negative values indicate relationships between nodes of different degree. In general, $r\in[-1;1]$. When $r=1$, the network is said to have perfect assortative mixing patterns, when $r=0$, the network is non-assortative, while, at $r=-1$, the network is completely disassortative.

Networks related to real social phenomena are \emph{assortative}, while networks associated with technical or biological systems are more often \emph{disassortative}~\cite{Newman2002,Regan2009}.

In graph theory and network science, centrality assigns numbers or rankings to nodes within a graph corresponding to their network position. Applications of this include identifying the most influential person(s) in a social network. The following types of centralities are used.

\emph{Betweenness centrality} is a measure of the centrality in a graph based on shortest paths~\cite{Freeman1977}. For every pair of nodes in a connected graph, there exists at least one shortest path between the nodes such that either the number of edges that the path passes through (for unweighted graphs) or the sum of the weights of the edges (for weighted graphs) is minimized. The betweenness centrality for each node is the number of these shortest paths that pass through the node. More concisely, the betweenness can be represented as~\cite{Brandes2001}:
\begin{equation}\label{eq:g}
g(k)= \sum_{i \neq k \neq j}\frac{\sigma_{ij}(k)}{\sigma_{ij}},
\end{equation}
where $\sigma_{ij}$ is equal to the total number of shortest paths from node $i$ to node $j$, and $\sigma_{ij}(k)$ is equal to the number of these paths passing through $k$.

In a connected graph, \emph{closeness centrality (or the closeness of a node)} is a measure of the centrality in a network, calculated as the reciprocal of the sum of the lengths of the shortest paths between the node and all other nodes in the graph. Thus, the more central a nodes is, the closer it is to all the other nodes~\cite{Sabidussi1966}. In our study, normalized closeness centrality is used~\cite{Freeman1978}.
\begin{equation}\label{eq:closeness_centrality}
C_i= \frac{N - 1}{\sum _{j}d_{j,i}},
\end{equation}
where $d_{j,i}$ is equal to the distance between the nodes $i$ and $j$, while $N$  is the number of nodes in the network.

\emph{Eigenvector centrality} determines that the centrality of each node is the sum of the centrality values of the nodes to which it is connected. The eigenvector centrality is determined by the eigenvector associated with the largest eigenvalue of the adjacency matrix $\mathbf{A}$. Formally,~\cite{Newman2008}
\begin{equation}\label{eq:eigencentrality}
x_i=\frac{1}{\lambda} \sum_{j}\mathbf{A}_{ij} x_j
\end{equation}
or, in matrix form,
\begin{equation}\label{eq:eigencentralitymatrix}
\mathbf{A}\mathbf{x} = {\lambda}\mathbf{x},
\end{equation}
where $\mathbf{x}$ is the senior right eigenvector, while $\lambda$ is the greatest eigenvalue.

Different centralities reflect different assumptions about what turns a character into one of the main characters~\cite{Everton2022}. Thus, focus on valency assumes that importance of a character depends on the number of direct links between this character and other characters. Focus on betweenness assumes that more important is the number of shortest path between characters passing through the character under consideration. For instance, a character owning only two links can be very important when it connects two large communities. Focus on closeness assumes that shortest path distances to all other characters are more valuable; the closer the better. Focus on eigenvector assumes that a character's importance depends on the significances of connected characters. (Recall idioms \emph{A man is known by the company he keeps} and\emph{ Birds of same feather flock together}.)

For simplicity, we present the normalized values of centralities in the tables.

\section{Results of the analysis}\label{sec:Results}

Let us briefly discuss the information on centralities values provided in Tables~\ref{tab:centralitiesmaster}, \ref{tab:yersh}, and \ref{tab:centrality} for the literary researcher. The degree (or valency, $\deg V$) determines any character's number of communication connections. Betweenness centrality ($g$) reflects the ability of a character to perform a mediating function and connect disparate communities communication through the formation of communication chains. Closeness centrality (or closeness, $C$) indicates, for each character, their location in a complex communication model, and characterizes the distances between network participants. Eigenvector centrality (also called eigenvector centrality or the prestige score, $x$) indicates the centrality or marginality of a character's communication partners. 

Table~\ref{tab:centralitiesmaster} presents the main characteristics of the main characters: degree, betweenness centrality, closeness centrality, and eigenvector centrality. For comparison, the last column of Tab.~\ref{tab:centralitiesmaster} presents the number of mentions of the character in the novel. Only a characters name in any form has been accounted for (e.g., Koroviev, Fagot, Fagot--Koroviev, Koroviev--Fagot), while other mentions of the same character have been ignored (e.g., assistant, interpreter, check-clad man, church choirmaster). The word counting was performed using The Russian National Corpus (\href{https://ruscorpora.ru}{ruscorpora.ru}).
\begin{table}[!htbp]
\tbl{Centrality table of the novel (main characters arranged by valency). Here, $g$ is the betweenness centrality~\eqref{eq:g}, $C$ is the closeness centrality~\eqref{eq:closeness_centrality}, and $x$ is the eigenvector centrality~\eqref{eq:eigencentrality}.\label{tab:centralitiesmaster}}
{\begin{tabular}{@{}rlccccc@{}} \toprule
&  Character &  $\deg V$ &  $g$ &  $C$ &  $x$ & Freq\\ \colrule
13 & Koroviev & 19 & 1.000 & 1.000 & 1.000 & 270 \\
38 & Margarita& 13 & 0.398 & 0.857 & 0.781 & 621 \\
2 & Bezdomny & 12 & 0.459 & 0.829 & 0.596 & 392 \\
4 & Woland & 11 & 0.904 & 0.990 & 0.828 & 265 \\
22 & Behemoth & 11 & 0.191 & 0.786 & 0.787 & 90 \\
23 & Azazello & 8 & 0.118 & 0.760 & 0.683 & 152 \\
28 & Bosoi & 9 & 0.397 & 0.605 & 0.186 & 135 \\
31 & Warenukha & 7 & 0.088 & 0.722 & 0.550 & 103 \\
37 & Master & 7 & 0.069 & 0.809 & 0.675 & 164 \\
9 & Levi Matthew & 3 & 0.595 & 0.643 & 0.124 & 5 \\
\botrule
\end{tabular}}
\end{table}

The research revealed characters with high centrality values, viz., Koroviev, Margarita, Bezdomny, Woland, Behemoth, Azazello, Bosoi, Warenukha, Master, and Levi Matthew. The highest ratings of Koroviev emphasize his function as an intermediary between the infernally fantastic retinue of Woland and the muscovites. Woland has fewer direct connections, viz., the valency (or degree, $\deg V$) of Koroviev is 19, while the valency of Woland is 11; however, the values of betweenness centrality ($g$), closeness centrality ($C$), and eigenvector centrality ($x$) of these characters are almost identical. Thus, Woland, located in the central community, serves as a junction maintaining connections with the main network participants, however, he restricts his contacts. This correlation emphasizes the dominant status of the character, who prefers to assign important meetings to his assistants. The fact that Woland and his entourage of trickster characters are placed at the novel's network centre of communicative interactions confirms, once again, the dominant role of such elements of the satirical novel in the heterogeneous genre structure of the text. Although Behemoth and Koroviev are described as an `inseparable couple', Koroviev is an undeniable leader of this couple, while Behemoth is only his assistant. There are several dialogues, in which Koroviev is involved, while Behemoth is not. This results in smaller values of the all centralities of Behemoth as compared to Koroviev. Higher values of all centralities of Koroviev as compared to other members of the Woland’s entourage are quite expected, since he is a former knight, which penance for his ill-timed joke was to turn into a joker for ages.

The characters network indicates the significance of the Margarita and Bezdomny; that corresponds to the formal and substantive features of the novel, viz., these characters are involved in a circle of motives (knowledge, testing, finding and renewal), forming the plot of `becoming' (one of the key plots in the novel). Margarita is one of the most communicative characters; however, her interlocutors belong mostly to the otherworld and to her closest social circle (the Master, her maid, and her neighbour). This results in large values of the valency, closeness, and eigenvector centrality, while the betweenness centrality demonstrates an intermediate value.

The Master is involved in communications with a small number of characters. In a significant number of chapters, he is not mentioned at all or he is mentioned but he does not present at a stage. He communicates with Margarita, Bezdomny, and infernal characters. As a result, his valency is intermediate ($\deg V = 7$), his betweenness is small ($g =0.069$),  while his closeness centrality value ($C = 0.809$) and eigenvector centrality value ($x=0.675$) both are above the average values. Since his interlocutors belong to the most influential characters, his closeness and eigenvector centralities are rather large. Tables~\ref{tab:centrality} and~\ref{tab:centralitiesmaster} demonstrate, that Master's centrality values are a little bit inconsistent with his position in the novel as a titular character. Master's communicative `passiveness' does not make him a minor or episodic character, he is at the centre of the characters network structure, showing his ability to maintain connections with characters occupying high positions in the network hierarchy.

Bezdomy connects several communities, joins the real world and otherworld. Although he is isolated in a hospital during almost the whole novel, his conversations with main characters both inside and outside the hospital leads to large values of the closeness centrality. However, his eigenvector centrality has intermediate value, since not only main characters are presented among his interlocutors.

Surprisingly, Warenukha belongs to main characters according to our network analysis. However, he is one of the characters, which connect other world and Moscow. He converses both with Woland and his encourage and with muscovites.

\begin{figure}[!htbp]
\centering
\includegraphics[width=\textwidth]{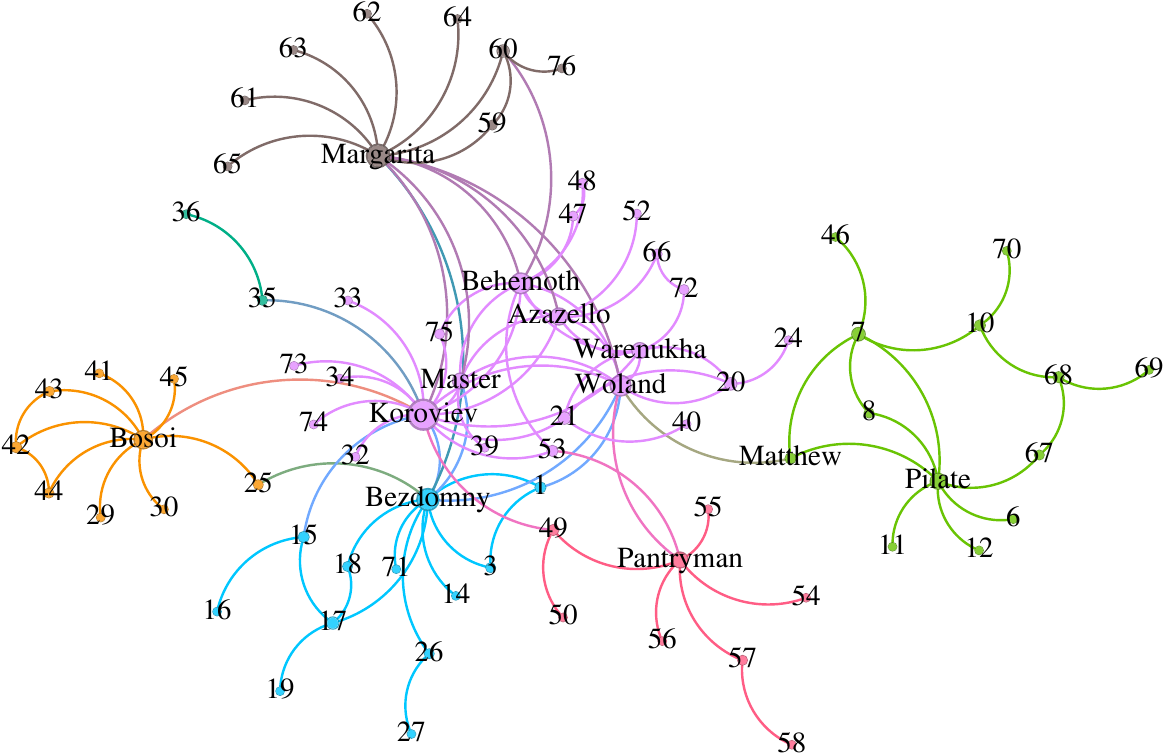}\\
\vspace*{8pt}
\caption{The largest connected component of the characters network (paired verbal communications) of the novel. Node numbers correspond to those in Table~\ref{tab:centrality}. Some important characters are indicated by their names. The node size is proportional to the valency of the node. Different communities are shown in different colours. The network image produced using Gephi~\cite{Bastian2009}.}\label{fig:finenetwork}
\end{figure}

The network of interactions between the characters is shown in Fig.~\ref{fig:finenetwork}. The assortativity coefficient is $-0.133$. Although the assortativity coefficient is negative, its magnitude is too small to consider this network as pronounce disassortative. The network may be treated as neither assortative nor disassortative. 

The structure of communities in the network was analysed using \texttt{igraph} library~\cite{igraph}, viz., the Girvan---Newman algorithm~\cite{Girvan2002} and the modularity~\cite{Newman2006}. Both algorithms resulted in the same partition of the network. Figure~\ref{fig:finenetwork} demonstrates the two main communities, viz., Yershalaim (smaller one) and Moscow (lager one). This division is clear, obvious, and expected. The Moscow community is divided in some subgroups. Each of these subgroups corresponds mainly to a particular location (hospital, theatre, MASSOLIT, etc.) and/or to a particular character, i.e., communities  are formed around Bezdomny (1), Woland (2), Bosoi (4), Sempleyarovs (5), Margarita (6), and the pantryman (7). The subnetwork of the Yershalaim part of the novel (3) stands out. This subnetwork is connected to the rest of the network only by the connection of Levi Matthew and Woland (dialogue on the roof). Here, the
number of the community corresponds to the last column in Table~\ref{tab:centrality}. All the communities are nonoverlapping. The modularity of the network is  $0.59$.

Figure~\ref{fig:finenetwork}, which shows the complete interaction network (paired verbal communications) of the characters together with the subdivision of the novel into communities, reflects the individual principles of plot structuring. Isolated vertex groups have structural similarities and schematically represent the branches that carve the novel's main storyline. The independence of the episodes is reflected in the network, where almost autonomous communities of characters operate, connected to the central community via a mediator character (Bezdomny's chase, Bosoi's dream, the pantryman's examination, and Margarita's flight). The community, which includes the \emph{Novel about Pilate}, has approximately the same number of vertices and edges as the other communities, and similar centrality values for its hanging vertices. The incidence of vertices 4 (Woland) and 9 (Levi Matthew) to one edge reflects disjunction of the closed space of the \emph{Novel about Pilate}, and reflects blurring of the `creative frame' between the external (Moscow) and the internal (Yershalaim) texts, which allows you to reconsider the plot composition. Perception of the novel as a `text within a text' can be adjusted by a further reading possibility, viz., the story of Pilate is a kind of prologue, presented in the form of chapters alternating with the novel's main action.

Separately, an analysis of the centralities was carried out in the Yershalaim part of the novel (the \emph{Novel about Pilate}). For this, the connection between Levi Matthew and Woland was broken, which meant the Yershalaim community was isolated. Quite consistent with the Master's statement that he is writing the \emph{Novel about Pilate}, all Pilate's centralities have a maximum value in this subnetwork. The values of Yeshua's centralities are the second largest (see Table~\ref{tab:yersh}).
\begin{table}[!htb]
\tbl{Centrality table (Yershalaim, characters are arranged by valency). Here, $g$ is the betweenness centrality~\eqref{eq:g}, $C$ is the closeness centrality~\eqref{eq:closeness_centrality}, and $x$ is the eigenvector centrality~\eqref{eq:eigencentrality}\label{tab:yersh}}
{\begin{tabular}{@{}lcccc@{}}
\toprule
Character &  $\deg V$ &  $g$ &  $C$ &  $x$\\ \colrule
Pontius Pilate & 7 &1.000&1.000& 1.000\\
Yeshua & 5 &0.699&0.885& 0.842\\
Judas of Kiriath  & 3 &0.466&0.577& 0.370\\
Niza & 3 &0.356&0.423& 0.254\\
Aphranius & 2 &0.274&0.577& 0.384\\
Mark Krysoboyi  & 2 &0.000&0.423& 0.563\\
Levi Matthew & 2 &0.000&0.423& 0.563\\
Secretary & 1 &0.000&0.231& 0.306\\
Legion Commander Legate  & 1 &0.000&0.231& 0.306\\
Joseph Kaifa  & 1 &0.000&0.231& 0.306\\
Executioner with a spear & 1 &0.000&0.192& 0.258\\
Juda's Killer  & 1 &0.000&0.077& 0.113\\
Niza's Maid  & 1 &0.000&0.000& 0.078\\
\botrule
\end{tabular}}
\end{table}

\section{Conclusion}\label{sec:Conclusion}
An analysis of the structure of characters network in the novel \emph{The Master and Margarita} by M.~A.~Bulgakov has been performed. Character interactions have been treated as paired verbal communications. Naturally, this approach is only one among a number of different approaches. However, the connections between the characters defined in such a way, provide a possibility to reduce subjectivity and to obtain results that are consistent with intuitive ideas. On the contrary, a broader interpretation of the interaction of characters leads to results that are difficult to interpret~\cite{Danilova2021arxiv}. Appendix presents some examples of problems, viz.,  (i) accounting for frontal communications, (ii) weights, and (iii) directional interactions. However, such a narrow understanding of the interaction may cut off some story lines. For instance, Vasily Stepanovich Lastochkin the accountant produces an isolated community, which contains himself, an unnamed investigator, an unnamed taxi driver, an unnamed young lady in the branch office of the Theatrical Commission, and an unnamed clerk. Obviously, this community is resulted by the black magic show in the Variety Theatre; however, a formal dialog-based study cannot demonstrate this. If other interactions would be taken into account, this community will be significantly larger.

Table~\ref{tab:centralitiesmaster} shows the characters in the novel that have the highest centrality. Table~\ref{tab:centralitiesmaster} indicates that the title of the novel does not correspond to the most influential characters, according to the network analysis. Neither the Master nor Maragritha are central characters in any of the measures of centrality. Despite the opinion of M.~A.~Bulgakov himself, that he was writing a novel about the devil, Woland is not the most influential character as determined by any of the measures of centrality; he shares first place with Koroviev only in terms of closeness. Only the character written about by the Master in the \emph{Novel about Pilate}, Levi Matthew, who connects the world of Yershalaim with the contemporary world of the author by his dialogue with Woland, belongs to the central characters, according to the network analysis. Thus, the emotional significance of a character is not necessarily related to its centralities in the network.

Dialog-based network analysis has evidenced that:
 \begin{enumerate}
\item there are two large communities (Yershalaim and Moscow), the latter divided into smaller communities;
\item  the central characters of the Yershalaim part of the novel are Ponius Pilate and Yeshua;
\item  according to our network analysis, Woland is not the central figure of the work, that is, the novel can hardly be called a novel about the devil. Thus, changing the original title of the manuscript of the novel to the final one looks quite reasonable.
\end{enumerate}

A distinctive feature of the network characters relationships in M.~Bulgakov's novel are the following:
\begin{enumerate}
  \item the division of the characters into groups that we have undertaken is partially reflected in the graph: only the characters of the \emph{Novel about Pilate} form a separate community;
  \item despite the framework composition of the text, the central community and the community of the \emph{Novel about Pilate} are connected by a mediator character;
  \item the formation of clusters is caused by micro-plots in which the main actors are characters who often occupy a secondary position in the novel plot.
\end{enumerate}

The Moscow and Yershalaim communities are separated in space and time. In the novel \emph{The Master and Margarita}, the connection between the members of these communities arises only through otherworldly forces, since, obviously, their real interaction is impossible. Beyond the analysis of dialogues, we can assume that the Master as an author is associated with each of his literary characters, which have real prototypes. However, this is precisely the connection with the literary characters of the \emph{Novel about Pilate}, and by no means with the historical figures who became the prototypes of the literary heroes of the \emph{Novel about Pilate}. Note that Woland insists on the historicity of the characters described by the Master. Margarita’s fascination with the work of her lover can hardly be seen as any kind of interaction with the characters of the \emph{Novel about Pilate}. The acquaintance of Berlioz and Bezdomny with the \emph{Gospel of Satan} can hardly be attributed to any kind of the interaction. Actually, in the Moscow community, there is only a material inanimate object, viz., the text of the novel, but not its characters. In fact, the characters network in \emph{The Master and Margarita} is multi-layered. However, the natural and obvious division of characters --- inhabitants of ancient Yershalaim, Moscow contemporaries of Bulgakov, and otherworld --- can hardly be reflected in the structure of the network with any way of choosing the rule for its construction (with any interpretation of interaction). Even the presence of some characters as two entities (human and otherworldly) hardly gives the key to construct a characters network in which these three worlds of the novel --- historical, contemporary, and otherworldly --- will be located in different layers of the characters network.


In the course of our study, we encountered a number of problems. The first was of how robust the results of network analysis were against random errors or imperfections in the method of taking into account the interactions of the characters. Comparison with preliminary analysis performed using extended interpretation of the relationships between characters~\cite{Danilova2021arxiv}, suggests that the way in which the relationships between characters are determined is critical. Thus, the extended network includes 153 characters

The second problem is determining a definition of interaction. Should frontal communications and monologues be considered as interactions of characters? When such communications are taken into account, the network becomes a directed graph. Should the direction of the links be taken into account? Should we take into account the intensity of communication, i.e., use a weighted graph? Also, should we treat knowledge rather than interaction as a social connection? Should silent or off-stage characters be considered when building a characters network? Should an indirect action be taken into account as an interaction, e.g., the destruction of Latunsky's apartment by Margarita? The action of Margarita is directed at Latunsky, although indirectly. The death of Berlioz is provoked by Annushka, although the connection between Annushka and Berlioz is also mediated. Appendix presents some examples of such the problems along with our comments.

Another issue is the problem of automating the study of character connections in works of art. If one adheres to the assumption that a connection between two characters (vertices) is created if two characters are mentioned in the same sentence or paragraph, then the characters network will be built incorrectly. For example, in Chapter~5 of  \emph{The Master and Margarita}, the members of MASSOLIT are obviously familiar with each other, but most of the characters are not mentioned together. In Chapter~13, the Master never mentions the name of his beloved, although the chapter is almost entirely devoted to her. Therefore, one can only build a proper characters network manually, which definitely complicates the work of analysing the structure of the network.

Network analysis of the verbal communication structure in \emph{The Master and Margarita} is of interest for modern literary studies as a digital method of data processing and visualization, used for evaluation of the text (studied thoroughly and deeply), providing  new perspectives for its research and paving the way for interdisciplinary interactions.

Quantitative analysis of verbal communications may be used as an auxiliary tool for interpreting Bulgakov's dramatized narrative, where dialogues become the main form of character contacts. Meanwhile, such analysis has a number of limitations, e.g., aspects of content intonation and emotional expressiveness of the communicants' messages, while the ideological value positions of both the characters and the author are also out scope of the observation. The results presented reflect only the relations-based parameters of verbal communication; they schematize it.

One of the main goals of network analysis is the identification of characters with high centrality values, allowing the most active subjects of action to be defined; however, these are not always important for explaining the author's artistic and philosophical intent.

Accordingly, network analysis enables accurate determination of the social connections circle of a particular character, quantifies them, mathematically describes the communicative activity and graphically visualizes that character's position in the structure of social connections, while schematically representing and correlating plot lines. All this is subsequent information to confirm the existing conclusions, and may also provide an impetus for further study of certain aspects of the work.

The prospects of the application of digital methods for data analysis and processing to study the organization of characters network in literary texts are determined by the necessity to study works that have common genre features or belong to the same aesthetic paradigm.

In particular, it is possible to make comparative analysis of the social relationships of Bulgakov's novel, with social ties to satiric works, united by trickster characters, viz., the \emph{Odesskie rasskazy} cycle (\emph{Odessa stories}) by I.~Babel, the novel \emph{Neobychajnye pohozhdeniya Hulio Hurenito} (\emph{Extraordinary adventures of Julio Hurenito}) by I.~Ehrenburg, dilogy by I.~Ilf and E.~Petrov (\emph{Dvenadtsat stul'ev} (\emph{Twelve chairs}) and \emph{Zolotoj telenok} (\emph{Golden calf})). It would also be interesting to use a network analysis to study V.~Nabokov's \emph{Dar} (\emph{Gift}), created almost at the same time as \emph{The Master and Margarita}, which implements the genre strategies both of a novel about an artist and the structure of a \emph{text within a text}.

\appendix

\section{Example of a verbal communication analysis}
Table~\ref{tab:example} presents an example of the extraction of the character communications. Hereafter quotations were taken from Mikhail Bulgakov \emph{The Master and Margarita} (Translated from the Russian by Michael Glenny. Published by Collins and Harvill Press, London, 1967). The spelling of names in quotes may differ from that used in our text.
\begin{table}[!htb]
\tbl{Example of a verbal communication analysis.\label{tab:example}}
{\begin{tabular}{@{}p{0.55\textwidth}p{0.3\textwidth}p{0.1\textwidth}@{}}
  \toprule
  Text & Characters & Network \\
  \colrule
  `A glass of lemonade\textsuperscript{a}, please,' said Berlioz. & Berlioz (1) $\rightarrow$ woman (3) &\\
`There isn't any,' replied the woman in the kiosk. For some reason the request seemed to offend her. & Berlioz (1) $\leftarrow$ woman (3) & \\
`Got any beer?' enquired Bezdomny in a hoarse voice. & Bezdomny (2) $\rightarrow$ woman (3) & \\
`Beer's being delivered later this evening' said the woman\textsuperscript{b}. & Bezdomny (2) $\leftarrow$ woman (3) &  $1 - 3 - 2 $\\
`Well what have you got?' asked Berlioz. & Berlioz (1) $\rightarrow$ woman (3) & \\
`Apricot juice, only it's warm' was the answer [woman].\textsuperscript{c} & Berlioz (1 ) ← woman (3) & \\
`All right, let's have some.' & Berlioz (1) $\rightarrow$ woman (3)& \\
  \botrule
\end{tabular}}
  \begin{tabnote}
Table notes
\end{tabnote}
\begin{tabfootnote}
\tabmark{a} In Russian original text `narzan' (a kind of the mineral water).\\
\tabmark{b} In Russian original text ‘replied’ [to Bezdomny].\\
\tabmark{c} In Russian original text the character is explicitly indicated as ‘said the woman’.
\end{tabfootnote}
\end{table}

\section{Complete character table}

\begin{longtable}[c]{rp{0.4\textwidth}ccccc}
\caption{Centrality table. Here, $g$ is the betweenness centrality~\eqref{eq:g}, $C$ is the closeness centrality~\eqref{eq:closeness_centrality}, and $x$ is the eigenvector centrality~\eqref{eq:eigencentrality}. The last column indicates membership in the community.\label{tab:centrality}}\\
\hline
&  Character &  $\deg V$ &  $g$ & $C$ &  $x $ & Cmty\\
\hline
\endfirsthead
\hline
\multicolumn{6}{c}{Table continuation}\\
\hline
&  Character&  $\deg V$ &  $g$ & $C$ &  $x $& Cmty\\
\hline
\endhead
\hline
\endfoot
\hline
\endlastfoot
1  & Berlioz   & 3 &0.010&0.577& 0.218 &1 \\
2  & Bezdomny & 12 &0.459&0.829& 0.596 &1 \\
3  & Beverage saleswoman & 2 &0.000&0.449& 0.115 &1 \\
4  & Woland & 11 &0.904&0.990& 0.828 &2 \\
5  & Pontius Pilate & 7 &0.301&0.403& 0.024 &3 \\
6  & Secretary & 1 &0.000&0.204& 0.003 &3 \\
7  & Yeshua & 5 &0.243&0.401& 0.023 &3 \\
8  & Mark Krysoboyi  & 2 &0.000&0.214& 0.007 &3 \\
9  & Levi Matthew & 3 &0.595&0.643& 0.124 &3 \\
10 & Judas of Kiriath & 3 &0.113&0.217& 0.003 &3 \\
11 & Legion Commander Legate & 1 &0.000&0.204& 0.003 &3 \\
12 & Joseph Kaifa & 1 &0.000&0.204& 0.003 &3 \\
13 & Koroviev & 19 &1.000&1.000& 1.000 &2 \\
14 & naked lady & 1 &0.000&0.446& 0.084 &1 \\
15 & Archibald Archibaldovich & 3 &0.076&0.561& 0.162 &1 \\
16 & Doorman & 1 &0.000&0.301& 0.023 &1 \\
17 & Ryukhin & 4 &0.066&0.472& 0.124 &1 \\
18 & Doctor & 2 &0.000&0.462& 0.102 &1 \\
19 & woman in clinic  & 1 &0.000&0.247& 0.018 &1 \\
20 & Likhodeev & 4 &0.068&0.561& 0.236 &2 \\
21 & Rimskyi& 4 &0.069&0.566& 0.258 &2 \\
22 & Behemoth & 11 &0.191&0.786& 0.787 &2 \\
23 & Azazello & 8 &0.118&0.760& 0.683 &2 \\
24 & smoker in Yalta & 1 &0.000&0.301& 0.033 &2 \\
25 & Praskovya Fedorovna & 2 &0.028&0.510& 0.111 &4 \\
26 & Dr. Stravinsky & 2 &0.058&0.454& 0.086 &1 \\
27 & Dr. Fedor Vasilievich  & 1 &0.000&0.235& 0.012 &1 \\
28 & Bosoi& 9 &0.397&0.605& 0.186 &4 \\
29 & Pelageya Antonovna & 1 &0.000&0.324& 0.026 &4 \\
30 & citizen1 & 1 &0.000&0.324& 0.026 &4 \\
31 & Varenukha & 7 &0.088&0.722& 0.550 &2 \\
32 & Bengalsky & 1 &0.000&0.531& 0.142 &2 \\
33 & Brunette & 1 &0.000&0.531& 0.142 &2 \\
34 & Sempleyarov & 1 &0.000&0.531& 0.142 &2 \\
35 & Sempleyarov's wife & 2 &0.058&0.538& 0.144 &5 \\
36 & Sempleyarov's relative & 1 &0.000&0.286& 0.020 &5 \\
37 & Master & 7 &0.069&0.809& 0.675 &2 \\
38 & Margarita & 13 &0.398&0.857& 0.781 &6 \\
39 & Editor & 1 &0.000&0.434& 0.096 &2 \\
40 & Taxi driver & 1 &0.000&0.304& 0.037 &2 \\
41 & a person from `other place' & 1 &0.000&0.324& 0.026 &4 \\
42 & actor in Bosoi's dream & 3 &0.000&0.332& 0.035 &4 \\
43 & Dunchil & 2 &0.000&0.329& 0.031 &4 \\
44 & Kanavkin & 2 &0.000&0.329& 0.031 &4 \\
45 & Cook & 1 &0.000&0.324& 0.026 &4 \\
46 & Executioner with a spear & 1 &0.000&0.204& 0.003 &3 \\
47 & Prokhor Petrovich & 2 &0.000&0.429& 0.130 &2 \\
48 & Anna Richardovna & 2 &0.000&0.429& 0.130 &2 \\
49 & Poplavsky & 3 &0.089&0.597& 0.173 &7 \\
50 & Pyatnazhko & 1 &0.000&0.319& 0.024 &7 \\
51 & Pantryman & 7 &0.300&0.599& 0.197 &7 \\
52 & Aloisy Mogarych & 1 &0.000&0.411& 0.097 &2 \\
53 & Hella & 3 &0.042&0.610& 0.281 &2 \\
54 & Lady with a green bag & 1 &0.000&0.321& 0.028 &7 \\
55 & Lady-Pharmacist & 1 &0.000&0.321& 0.028 &7 \\
56 & Receptionist & 1 &0.000&0.321& 0.028 &7 \\
57 & Prof Kuzmin & 2 &0.058&0.329& 0.028 &7 \\
58 & Ksenia Nikitishna & 1 &0.000&0.158& 0.004 &7 \\
59 & Natasha & 2 &0.000&0.469& 0.146 &6 \\
60 &  Nikolai Ivanovich & 4 &0.060&0.500& 0.247 &6 \\
61 & Boy & 1 &0.000&0.462& 0.110 &6 \\
62 & A fat man& 1 &0.000&0.462& 0.110 &6 \\
63 & Goat-footed& 1 &0.000&0.462& 0.110 &6 \\
64 & Tofana & 1 &0.000&0.462& 0.110 &6 \\
65 & Frieda & 1 &0.000&0.462& 0.110 &6 \\
66 & Annushka & 2 &0.002&0.416& 0.110 &2 \\
67 & Aphranius & 2 &0.057&0.217& 0.003 &3 \\
68 & Niza & 3 &0.060&0.097& 0.001 &3 \\
69 & Niza's Maid & 1 &0.000&0.000& 0.000 &3 \\
70 & Juda's Killer & 1 &0.000&0.084& 0.000 &3 \\
71 & Investigator in clinic & 1 &0.000&0.446& 0.084 &1 \\
72 & Investigator & 2 &0.001&0.395& 0.093 &2 \\
73 & Doorman& 1 &0.000&0.531& 0.142 &2 \\
74 & saleswoman in a candy store dept & 1 &0.000&0.531& 0.142 &2 \\
75 & Sophia Pavlovna & 2 &0.000&0.554& 0.253 &2 \\
76 & Nikolai Ivanovich's spouse  & 1 &0.000&0.265& 0.035 &6 \\
%
\end{longtable}

\begin{figure}[!htbp]
\centering
\includegraphics[width=0.5\textwidth]{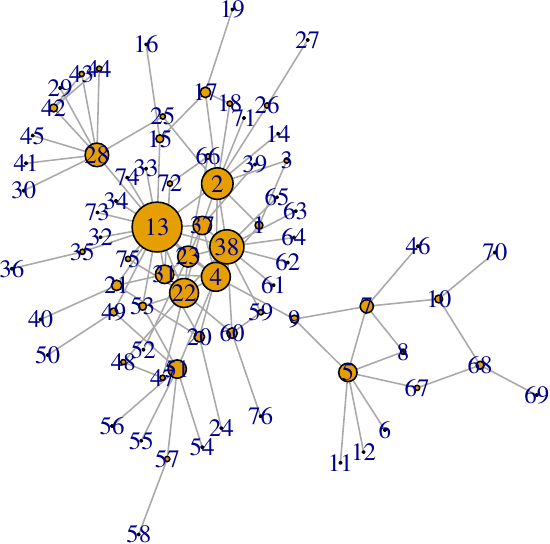}\\
\vspace*{8pt}
\caption{The complete network of interactions (paired verbal communications) of the characters of the novel. Node numbers correspond to those in Table~\ref{tab:centrality}.}\label{fig:network}
\end{figure}

The network of interactions between the characters is shown in Fig.~\ref{fig:network}.

\section{Some examples of frontal communications}
Some examples of frontal communications explaining why frontal communications are excluded from consideration. Some problems associated with the use of a directed graph (directed links) and weights are as well outlined.
\paragraph{Bidirectional frontal partially personalized communication within a limited personalized group.}

`There he ordered his waiting secretary to call the Legate of the Legion and the Tribune of the cohort into the garden, also the two members of the Sanhedrin and the captain of the temple guard, who were standing grouped round the fountain on the lower terrace awaiting his call.\\
$\langle \dots\rangle$

There in the presence of all the men he had asked to see, the Procurator solemnly and curtly repeated that he confirmed the sentence of death on Yeshua Ha-Notsri and enquired officially of the Sanhedrin members as to which of the prisoners it had pleased them to release. On being told that it was Bar-Abba, the Procurator said: \dots'

It is reported what Pilate said and what answer he received, but the author of the answer is not named. Thus, it is possible to build connections directed from Pilate to each of the members of the Sanhedrin, but the connection directed to Pilate does not have a definite beginning (a member of the Sanhedrin). The integration of the members of the Sanhedrin into a single group character is impossible due to the lengthy, preliminary private dialogue between Pilate and one of the members of the Sanhedrin, viz., Kaifa (Caiaphas). One could use Kaifa and the group character, i.e., the other members of the Sanhedrin, and then assume that it was Kaifa who gave the answer to Pilate, however, here we move from a solid base of facts to a shaky ground of assumptions, which we consistently tried to avoid.

\paragraph{Mass personalized frontal communication within a limited personalized group.}

Each speaker knows each listener, but does not address a specific listener;  in each case, the speaker is called by name, e.g., a meeting of the board of MOSSOLIT (Chapter 5). In this case, it would be appropriate to use weights that are inversely proportional to the number of board members, however, such a choice of weights is not possible with other frontal communications and contradicts the (highly controversial) idea of using dialogue size as weights.

\paragraph{Frontal communication with a limited, but not personalized group.}

For instance, a performance in Variety (addresses to the auditorium of Bengalsky, Fagot, Gela and separate replies from non-personalized spectators).

``Well, ladies and gentlemen, shall we forgive him?' asked Faggot, turning to the audience.

`Yes, forgive him, forgive him!' The cries came at first from a few individual voices, mostly women, then merged into a chorus with the men.'

\paragraph{Frontal communication with partially personalized listeners and a depersonalized response.} For example, Pilate's speech in the square, directed primarily to the crowd, but some of the listeners are personally known to him, for example, Yeshua. The crowd reacts to Pilate's speech, that is, formally, it can be considered as a bidirectional communication. The choice of weights in this case is also very problematic, as well as the choice of the beginning of the connection (the collective character `crowd'?).

\paragraph{Frontal communication with an indefinite and non-personalized group.}

`As the squadron commander, a Syrian as small as a boy and as dark as a mulatto, trotted past Pilate he gave a high-pitched cry and drew his sword from its scabbard.'

It clearly does not follow from the text that he addressed Pilate specifically. It can be assumed, for example, that he gave his subordinates the command `Attention! Alignment to the procurator!' or cried `Glory to the emperor!' Again, we move from a solid base of facts to the shaky ground of assumptions.

\paragraph{Frontal communication with a personified group that does not know or see the speaker.}

`She [invisible Margarita] saw a kitchen. Two Primuses were roaring away on a marble ledge\footnote{In  Russian original text 'stove'.}, attended by two women standing with spoons in their hands and swearing at each other.

`You should put the light out when you come out of the lavatory, I've told you before, Pelagea Petrovna,'
said the woman with a saucepan of some steaming decoction,' otherwise we'll have you chucked out of here.'

`You can't talk,' replied the other.

`You're both as bad as each other,' said Margarita clearly, leaning over the windowsill into the kitchen.

The two quarrelling women stopped at the sound of her voice and stood petrified, clutching their dirty spoons. Margarita carefully stretched out her arm between them and turned off both primuses. The women gasped. But Margarita was already bored with this prank and had flown out again into the street.'

\paragraph{Frontal communication, when all the interlocutors are personified, but the speaker's speech is addressed mainly to one of them.} The Woland's story about Pilate is a direct response to the Berlioz's remark. However, Bezdomny is another listener.

`But one must have some proof\dots ' began Berlioz.

`There's no need for any proof,' answered the professor. In a low voice, his foreign accent vanishing altogether, he began: `It's very simple--early in the morning on the fourteenth of the spring month of Nisan the Procurator of Judaea, Pontius Pilate, in a white cloak lined with blood-red\dots '

\section{Some issues concerning the weights}
Some examples demonstrating that the duration of communication does not always correspond to the importance of the connection between the characters who have entered into communication, therefore, can hardly be used as a weight.

\begin{enumerate}
  \item A few words of Margarita determine the future fate of Frieda.
  `The door burst open and a naked, dishevelled but completely sober woman with ecstatic eyes ran into the room and stretched out her arms towards Margarita, who said majestically :

`You are forgiven. You will never be given the handkerchief again.'

Frieda gave a shriek and fell spreadeagled, face downward on the floor in front of Margarita. Woland waved his hand and Frieda vanished.'

  \item A few words of the Master determine the further fate of Pilate.

  `The master seemed to be expecting this while he had been standing motionless, watching the seated Procurator. He cupped his hands to a trumpet and shouted with such force that the echo sprang back at him from the bare, treeless hills :

`You are free! Free! He is waiting for you!'

  \item Almost the entire first chapter and the entire second chapter are Woland's conversation with Berlioz and Bezdomny (the story about Pilate is addressed specifically to them). The summary dialogues of the Master and Margarita are significantly smaller than the second chapter. The same can be said about chapter \emph{Enter the Hero}: the total length of the Master's text addressed to Bezdomny is significantly longer than all his dialogues with Margarita. Thus, the weight of the link, proportional to the length of the speech and taking into account the direction of the speech, will result in the protagonist being Bezdomny. It is through speeches addressed to Bezdomny that the author introduces readers to an essential part of the history of Pilate and the history of the Master. It is clear that this way of presenting information is an exclusively literary device that does not reflect real social ties and their significance. (In the case of the Master's story, it is appropriate to recall the \emph{Stranger-on-the-train phenomenon} well-known to psychologists. Bezdomny becomes a random and one-time interlocutor of the Master due to a random coincidence.)
  \item Archibald Archibaldovich's communication with Koroviev and Behemoth in Griboyedov is informational noise, its duration can hardly characterize the importance of the connection between these characters.

      `The old stained tablecloth vanished instantly from the table and another, whiter than a bedouin's burnous, flashed through the air in a crackle of starch as Archibald Archibaldovich whispered, softly, but most expressively, into Koroviev's ear:

`What can I offer you? I've a rather special fillet of smoked sturgeon\dots I managed to save it from the architectural congress banquet\dots'

`Er\dots just bring us some hors d'oeuvres\dots' boomed Koroviev patronisingly, sprawling in his chair.

`Of course,' replied Archibald Archibaldovich, closing his eyes in exquisite comprehension.

$\langle\dots\rangle$

`A little breast of grouse, perhaps?' said Archibald Archibaldovich in a musical purr. The guest in the shaky pince-nez thoroughly approved the pirate captain's suggestion and beamed at him through his one useless lens.

$\langle\dots\rangle$

`Please excuse me --- I must go and attend to the grouse!'

  \item The rather lengthy communication between Koroviev and Sofia Pavlovna at the entrance to Griboyedov (one of the most frequently cited scenes in the novel) hardly allows us to attribute much weight to this connection, proportional to the length of the corresponding text: although this communication is very important, since it touches upon fundamental worldview issues, Sofia Pavlovna is episodic character.

  \item An example demonstrating that the mention of characters in one paragraph cannot always be considered as an explicit interaction between characters. In the quotations below, it is important that Margarita is invisible. The porter does not know who performs the described actions. Although Margarita can see the porter, his behaviour does not affect Margarita's actions in any way.

      `Margarita frowned at the inscription, wondering what the word `Dramlit' could mean. Tucking her broomstick under her arm, Margarita pushed open the front door, to the amazement of the porter, walked in and saw a huge black notice-board that listed the names and flat numbers of all the residents. The inscription over the name-board, reading `Drama and Literature House,' made Margarita give a suppressed yelp of predatory anticipation. Rising a little in the air, she began eagerly to read the names: Khustov, Dvubratsky, Quant, Beskudnikov, Latunsky\dots

`Latunsky!' yelped Margarita. `Latunsky! He's the man\dots who ruined the master!'

The porter jumped up in astonishment and stared at the name-board, wondering why it had suddenly given a shriek.

$\langle\dots\rangle$

Having dealt with all Latunsky's windows, Margarita floated on towards the next flat. The blows became more frequent, the street resounded with bangs and tinkles. The porter ran out of the front door, looked up, hesitated for a moment in amazement, popped a whistle into his mouth and blew like a maniac. The noise inspired Margarita to even more violent action on the eighth-floor windows and then to drop down a storey and to start work on the seventh.

Bored by his idle job of hanging around the entrance hall, the porter put all his pent-up energy into blowing his whistle, playing a woodwind obbligato in time to Margarita's enthusiastic percussion. In the intervals as she moved from window to window, he drew breath and then blew an ear-splitting blast from distended cheeks at each stroke of Margarita's hammer. Their combined efforts produced the most impressive results.\footnote{In Russian original text ‘His efforts, combined with those of the enraged Margarita, yielded great results.’} Panic broke out in Dramlit House.'

 \item An example demonstrating that the mention of characters even in one sentence cannot sometimes be treated that the characters know one another.

     'Glukharyov danced away with the poetess Tamara Polumesyatz. Quant danced, Zhukopov the novelist seized a film actress in a yellow dress and danced. They all danced--Dragunsky and Cherdakchi danced, little Deniskin danced with the gigantic Bo'sun George and the beautiful girl architect Semeikin-Hall was grabbed by a stranger in white straw-cloth trousers. Members and guests, from Moscow and from out of town, they all danced--the writer Johann from Kronstadt, a producer called Vitya Kuftik from Rostov with lilac-coloured eczema all over his face, the leading lights of the poetry section of MASSOLIT-- Pavianov, Bogokhulsky, Sladky, Shpichkin and Adelfina Buzdyak, young men of unknown occupation with cropped hair and shoulders padded with cotton wool, an old, old man with a chive sticking out of his beard danced with a thin, anaemic girl in an orange silk dress.'

     There is no reason to believe that the characters dancing in different pairs interact in any way with each other or even know each other.

\end{enumerate}

\section{Some issues concerning to the number of mentions in the text.}
\begin{enumerate}
  \item A character can be mentioned in differen manners, e.g., Woland is mentioned as Woland (265), messire (65), foreigner (62), professor (46), magician (35),  consultant (23), unknown man (9), Satan (8), lord (4), maitre (2), spirit of evil (2); Koroviev  is mentioned as Koroviev (227) and Fagot (46) including Koroviev-Fagot (2) and Fagot-Koroviev (1), church choirmaster (23), interpreter (19)), check-clad man (14), assistant (9, knight (6), buffoon (3).
  \item Some names (unknown man (stranger), professor, foreigner, knight, cat, wife, assistant, etc.) are not unique, i.e., their are used for different characters.
  \item Pronoun (she) is widely used in chapter \emph{Enter the Hero} when Master tell his love story.
\end{enumerate}
Thus, the number of mentions in the text can be either completely formal based on word frequency counting or informal based on a semantic analysis of word usage in context. In fact, the latter means an accounting for any interaction in a broad sense.

\bibliographystyle{ws-acs}
\bibliography{myth2}

\end{document}